\documentclass{article}
\usepackage{frascatiphys}
\begin{document}
\title{ Light Meson Spectroscopy from Charm Decays}
\author{ Jussara M. de Miranda \\
{\em Centro Brasileiro de Pesquisas Fisicas- CBPF} \\
{\em Xavier Sigaud, 150 - Rio de Janeiro, Brazil}
}
\maketitle
\baselineskip=11.6pt
\begin{abstract}

  We discuss recent achievements in light scalar mesons spectroscopy 
through amplitude analysis of charm particle decay and its consequences. The 
high  statistics clean  samples of charmed mesons,  in addition to it's
definite $ J^P $ and  mass, is turning these 
decays into a new important environment to study light meson physics.
We give special attention to the scalar sector favored by a high coupling 
to charm. 

\end{abstract}
\baselineskip=14pt
\section{Introduction}
Most mesons are well understood in the context of the quark model. This is not
true just for the scalar sector, that for long has been a source of
controversy. For the latest 30 years a great deal of experimental effort have
been made but in many cases the experimental results do not
converge to compatible outputs. In production experiments, observing the scalar
resonant states is difficult due to a large contribution of the non-resonant
background. Moreover, light scalars are too numerous within a relatively
short mass interval. The absence of a unique analysis procedure, especially
regarding the interference of the background with the resonances results in
conflicting measurements. In this sense the wider and lighter the state the worse,
and this is the case of the two states $\sigma(500)$ and $\kappa$ that will be
discussed here.

The theoretical interpretation of the lightest scalar meson have also been
unclear. Many  objects -- glueballs, $KK$ molecules, multiquark compact
 states -- are expected to populate the area. In a recent review article
 F. Close and N. A. T$\ddot{\rm o}$rnqvist\cite{ct} discuss
the scalars from both experimental end theoretical points of view. 
They suggest that the scalars be organized not in one but two nonets. 
The ``standard'' quark model $q{\bar q}$ nonet, distorted by a glueball
predicted by lattice QCD, is enough to explain  the region above $\sim$ 1 GeV. 
This nonet is composed by $a_0 (\sim 1400), f_0(1370), K(1430),
f_0(1500) $ and $f_0(1710)$. The states below $\sim$ 1GeV -- $f_0(980), a_0(980)$
and possibly $\sigma(500)$ and  $\kappa$ -- by arguments based on QCD
attractive forces in S-wave, would also form a nonet. The interpretation 
of such nonet would be more complex, of the type meson-meson. 
In the article, they stress the importance  of  charm decay as ``opening up a
new experimental window for understanding light meson spectroscopy and specially
the controversial scalar meson which are copiously produced in these decays''.

The use of charm decay is an alternative to the traditional production experiments to
study lighter resonant particles. It was made possible by the large clean
samples of charm now available and which are attributed to mankind effort but
also, and more important, by what can be seen as nature's
gifts:

1) Non-leptonic charm decays are preferentially two body or quasi-two body, through the formation of
intermediate resonant states. With a small non-resonant (NR) component one
avoids having to deal with model dependent amplitudes and its interference with
the resonant amplitudes. This feature is particularly important for the wide scalars
because their amplitudes and the NR (usually taken as constant) can become very
similar. We return to this issue later in the text.

2) Charm  couple strongly with scalars. This empirical fact is
  present in the decays discussed here where  the
contribution to scalars dominates all processes.

3) We are dealing with a well defined initial state; (charm meson mass, $M$ and
$J^P$)

The first light resonant parameters extracted from charm data use D meson  decays to three charged
pseudo-scalars.  In the following session we summarize the Dalitz plot amplitude formalism 
used. The parameterization of the overall amplitude consists of a coherent sum
of all contributions (NR flat and resonant as relativistic Breit-Wigners modulated by 
angular momentum conservation functions and form factors) with complex
coefficients obtained from
the fit. The magnitude of the coefficients are proportional to each  relative
contribution and the phases accommodate in an effective way the 
final state interactions. It should be  pointed out as the a great pro 
of these analysis
 the very good description that such simple model give to the data. 
 Alternative descriptions for the decay amplitude are
being tried, for example Focus experiment is using  K-matrix to parameterize
the light scalars contribution, but we shall not discuss those, still
preliminary,
results. Also expected for the near future are Focus  four-body amplitude analysis.

\section{Three body hadronic decay formalism}
We  describe here the analysis procedure used by the E791 collaboration
in their $D^+$ and $D^+_s \to \pi^+\pi^-\pi^+$\cite{ds}\cite{dp} and 
$D^+ \to K^-\pi^+\pi^+$ \cite{kpipi},
from which was measured masses and widths of $\sigma(500), f_0(980), f_0(1370)$,
 $\kappa$ and $K^*_0(1430)$.

The decay of a scalar hadron of mass $M$ into 3 spinless daughter particles
is completely specified with two degrees os freedom, conveniently 
chosen as two Dalitz plot variables, $m_{12}^2$ and $m_{23}^2$. The Dalitz
plot density distribution is proportional to the invariant decay amplitude
$\cal{A}$  squared and reflects the dynamics of the decay process.
A simple  analytical model for $\cal{A}$ is given by a coherent of all
intermediate states contributing, resonant or not: 

\begin{equation}
{\cal A} = a_{NR} e^{i\delta_{NR}} {\cal A}_{NR} + 
\sum_{j=1}^n a_j e^{i\delta_j}{\cal A}_j 
\label{e1}
\end{equation}

The parameters $a$ give the various relative contributions and the phases
$\delta$ are accounts for  final state interactions. 
The non-resonant amplitude ${\cal A}_{NR}$ is represented by a constant. Which
is a reasonable assuption if we imagine that it is dominantly $S$-wave, in any 
case, the   impact of this choice in the results is reduced due
to a small NR contribution observed. 
The decay through resonant intermediate states, $j$ (that then decay to the
observed $k$ and $l$ hadrons),  are viewed as s-channel 
processes where the
resonance plays the role of massive propagators, represented by relativistic
Breit-Wigner functions, $ BW_j$ \footnote{
 For the $f_0(980) \pi^+$ E791 uses a coupled channel Breit-Wigner, following
the parameterization used by  the WA76 Collaboration \cite{wa76},

$$
BW_{f_0(980)} = {1 \over {m_{kl}^2 - m^2_0 + im_0(\Gamma_{\pi}+\Gamma_K)}},{\rm
with}
$$

$$
\Gamma_{\pi} = g_{\pi}\sqrt{m_{kl}^2/ 4 - m_{\pi}^2}, {\rm and}
$$

$$
\Gamma_K = {g_K \over 2}\ \left( \sqrt{m_{kl}^2/ 4 - m_{K^+}^2}+
\sqrt{m_{kl}^2/ 4 - m_{K^0}^2}\right)
$$
} 
. Momentum dependent 
form factors, $F_D$ and $F_R$ describe the non-pointlike nature of the $D$ meson
and the resonance respectively and depend on the resonance spin, $J$ and the
radii of the relevant mesons.
 The angular momentum conservation is taken care by the function ${\cal M}^J_j$. Each
 resonant amplitude, ${\cal A}_j$ is written as:
 
\begin{equation}
{\cal A}_j = BW_j \times F_D \times F_R \times {\cal M}_j^J
\label{e2}
\end{equation}

\begin{equation}
 BW_j =  {1 \over {m_{kl}^2 - m^2_0 + im_0\Gamma_j(m_{kl})}}
\label{e3}
 \end{equation}

with

\begin{equation}
\Gamma(m_{kl}) = \Gamma_0 \frac{m_0}{m_{kl}}\left(\frac{p^*}{p^*_0}\right)^{2J+1}
\frac{F_R^2(p^*)}{F_R^2(p^*_0)}
\end{equation}
\noindent
Above  $m^2_{kl}$ is the the Dalitz plot variable, i.e. the 
invariant mass of the two hadrons forming a
spin-J resonance. Detailed expression of all the above functions are found in
the references \cite{ds,dp,kpipi}. Combinatorics background, detector 
efficiency and Bose symmetrization are also considered in a
maximum likelihood fit to the data to extract the parameters.
In most cases the masses, $m_0$, and widths, $\Gamma_0$ of the resonances 
are fixed by values listed in PDG. Only for new or poorly measure states they
are allowed to float in the fit.

\begin{figure}[t]
 \vspace{9.0cm}
\includegraphics{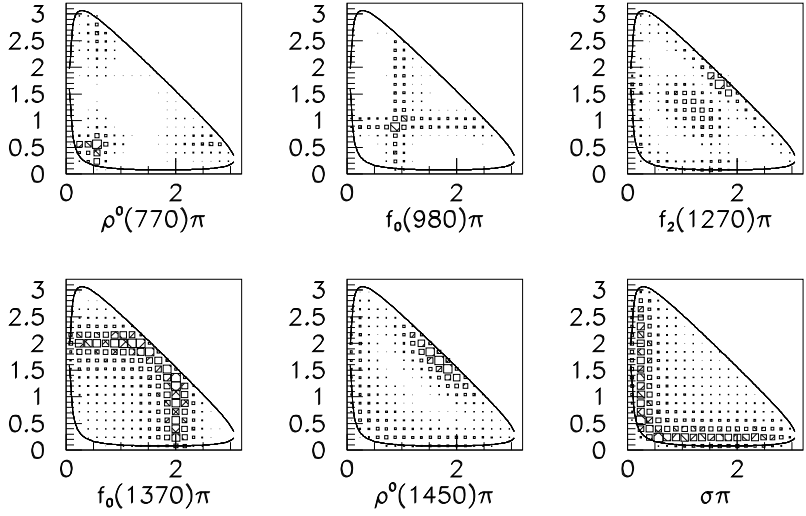}
 \caption{\it Dalitz-plot of the individual resonant contributions 
to the $D^+\to \pi^-\pi^+\pi^+$ decay.
    \label{fig1} }
\end{figure}

Figure \ref{fig1}  shows , as an example, Monte Carlo simulation 
Dalitz-plot of each individual
 resonant state
that contributes in the $D^+\to \pi^-\pi^+\pi^+$ decay. Notice that being 
 coefficients and the individual amplitudes 
 complex quantities, interference effects will take place when all the
pieces act together.

The above described produces a quite stringent model and obtaining acceptable
fits is usually a difficult task. 
To access the quality of each fit  a fast Monte Carlo (MC) program was developed
which produces  Dalitz plot event densities accounting for signal and background
PDF's, including detector efficiency and resolution. Comparing the MC density
distribution generated using parameters extracted from a  given 
fit  with that for the data, it is produced a $\chi^2$ distribution. When
comparing two possible models, the best discriminating power
test requires ensembles of Monte Carlo ``experiments''. In the $\kappa$ discussion below we illustrate the
technique.

\section{The $f_0'$s resonances}

\begin{figure}[t]
 \vspace{8.0cm}
\includegraphics{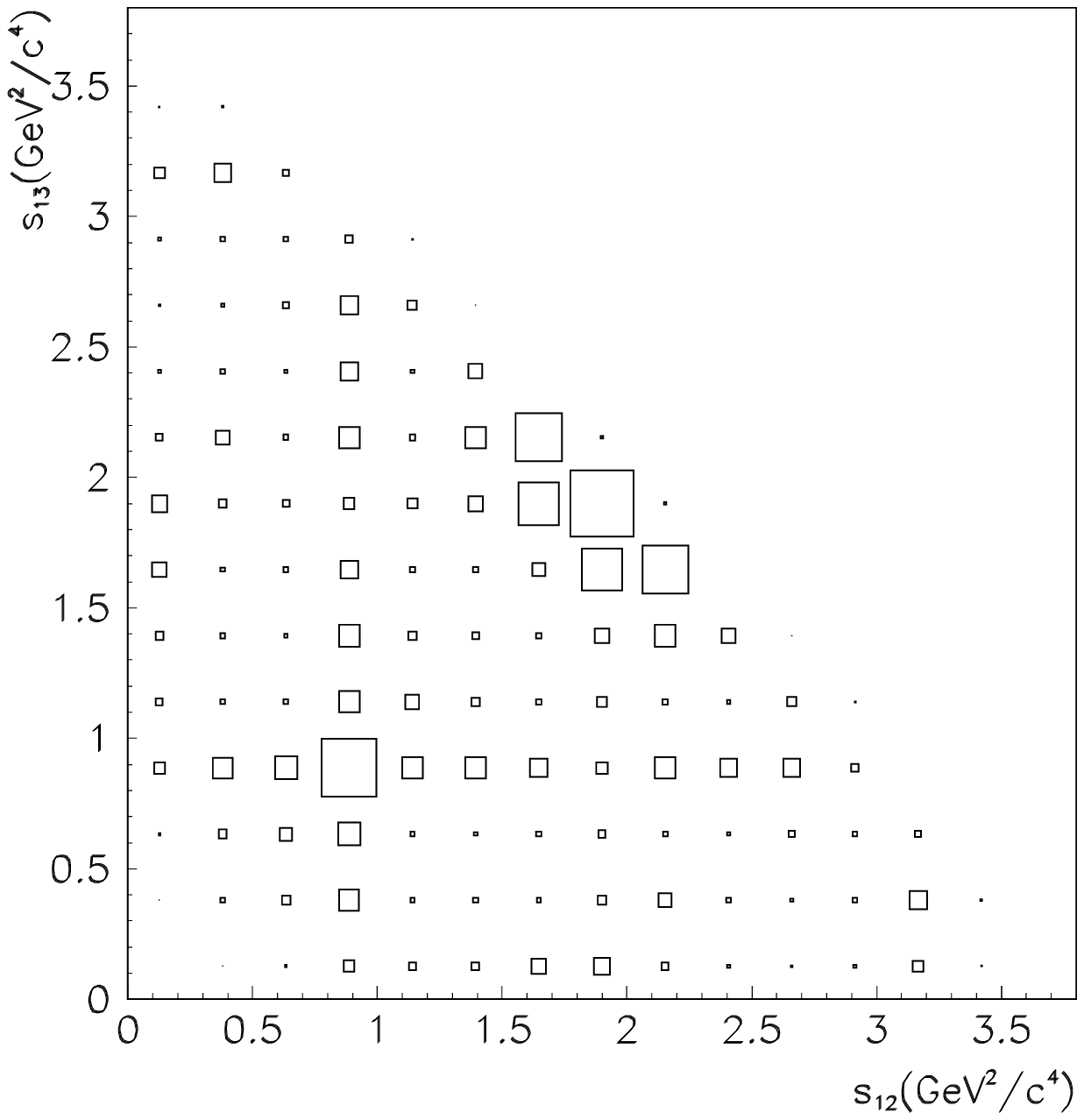}
\includegraphics{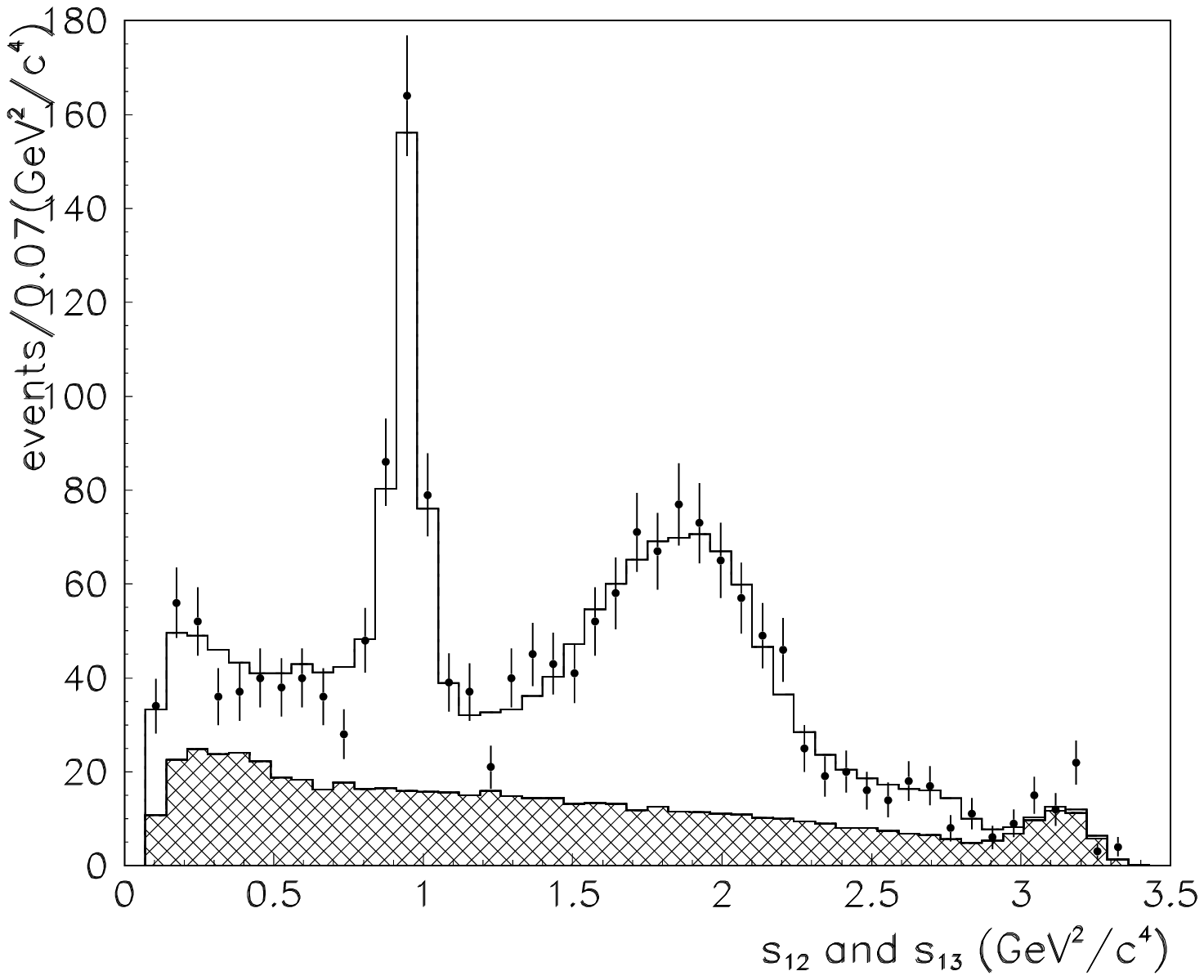}
 \caption{\it E791 $ D_s^+ \to \pi^+\pi-\pi^+$ Dalitz plot and 
$m^2_{\pi^+\pi^-}$  projection
    \label{fig2} }
\end{figure}

We have chosen to start the results sessions with the $f_0$ because of the extremely
clear signal seen in the $D_s^+ \to \pi^+\pi-\pi^+$ 
decay, figure \ref{fig2}.
There are dozens of measurements listed in PDG\cite{pdg} for this state and they converge to a
reasonably well defined mass, $980\pm 10$ MeV but the estimation for width is
from 40 to 100 MeV. 

For the best E791 fit  5 resonant channels contributes significantly:
$\rho^0(770)\pi^+$, $\rho^0(1450)\pi^+$, $f_0(980)\pi^+$,
 $f_2(1270)\pi^+$,  and  $f_0(1370)\pi^+$,
 plus a NR that contributed with a fraction of only $0.5\pm 0.2 $\%. The fit
  have $\chi^2/dof$ = 71.8/68 with a confidence level of 35\%. The
 dominant contributions comes from  $f_0(980)\pi^+$, $56.5\pm6.4$\%, 
 and $f_0(1370)\pi^+$, 32.4$\pm$7.9\% from which they measure:
 $m_{f_0(980)} = 977 \pm 3.6 $ MeV/c$^2,  g_\pi = 0.09\pm 0.01, g_K = 0.02\pm 0.05,
  m_{f_0(1370)}= 1434\pm 20$ MeV/c$^2$ and $\Gamma_{f_0(1370)} = 172 \pm
  33$ MeV/c$^2$.
  There is no evidence of a third higher mass scalar state, $f_0(1500)$.

\begin{figure}[t]
 \vspace{8.0cm}
\includegraphics{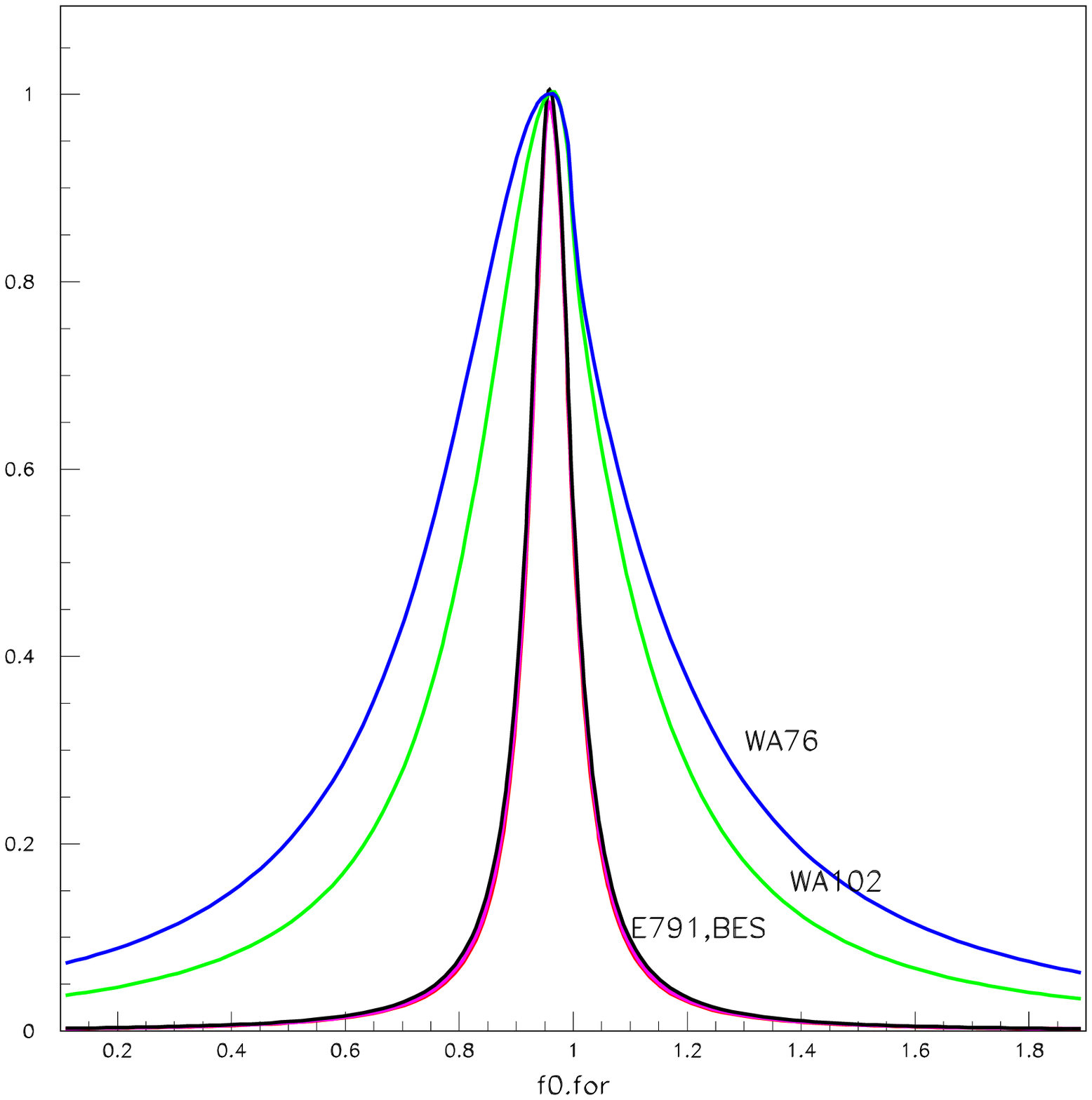}
 \caption{\it Comparison of several $f_0(980)$ results.
    \label{fig3} }
\end{figure}

Previous production experiments  claim a large contribution
of the $K{\bar K}$ channel by estimation a large value of the parameter $g_K$
\cite{wa76}\cite{wa102}.
These results do not agree with the value measured by E791 that, in fact,
finds that a simple Breit-Wigner \ref{e3} is sufficient to represent the
data. From the small but clear preliminary signal BES 
collaboration measure $m_{f_0} = 0.980\pm 0.009$ GeV and $\Gamma_{f_0}=
0.045\pm 0.30$ GeV\cite{bes1}. In figure \ref{fig3} we compare various results and
conclude that in charm decay the resonance $f_0(980)$ presents a narrower
signal.

\section{The $\sigma(500)$}
The 1999 the Workshop on Hadron Spectroscopy \cite{he99} devoted one entire session to the meson
$\sigma(500)$, `` What do we know about the $\sigma$?''. By that time E791 had
not published their observation of the meson $\sigma$ in the decay of 
$D^+ \to \pi^+\pi^-\pi^+$\cite{dp}. Today charm decay is viewed as a tool for studying
the light spectroscopy. 

\begin{figure}[t]
 \vspace{9.0cm}
\includegraphics{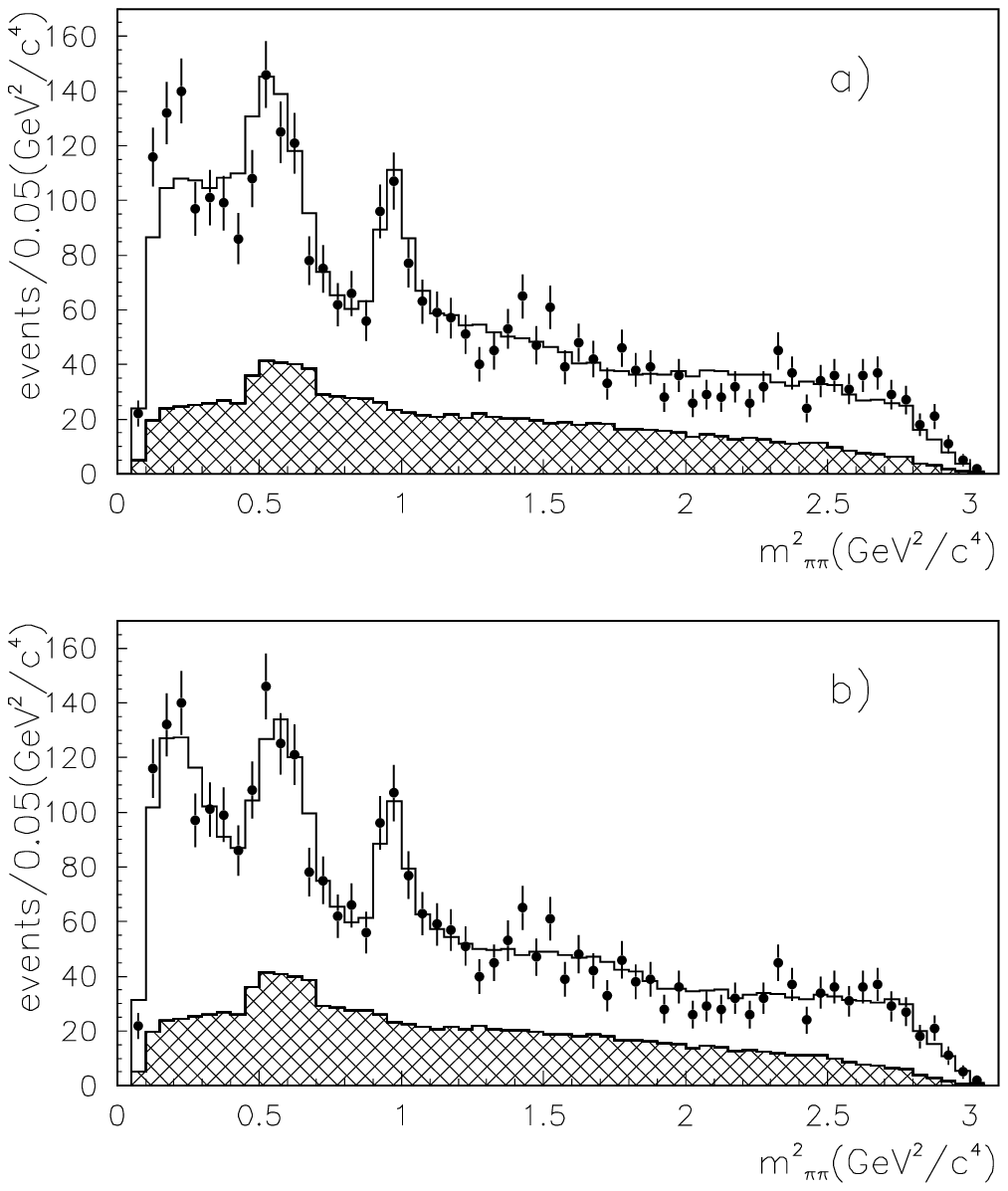}
 \caption{\it E791 $ D^+ \to \pi^+\pi-\pi^+$  
$m^2_{\pi^+\pi^-}$  projections for data  (error bars) and model (solid line).
Shaded area is the background. a) solution for Fit 1, and b) Fit 2.
    \label{fig4} }
\end{figure}

In figure \ref{fig4} we show the projection of the E791 Dalitz plot where a
clear peak at low mass can be seen. The figure compares the best fit achieved
with all possible well established resonances available at the time, $a)$ (Fit
1), to their
solution including a low mass scalar state, $b)$ (Fit 2). They measured $m_{\sigma} =
478\pm 29$ MeV/c$^2$ and $\Gamma_{\sigma} = 324\pm 46$ MeV/c$^2$ and the
inclusion of the state took them from an unacceptable solution of  
 $\chi^2/dof$ = 254/162 with a confidence level less than 10$^{-5}$ to a very
 good fit with $\chi^2/ dof$ = 138/162 and confidence level of 90\%. In Fit 1
 the NR  contribution is dominant with 38\% and in Fit 2 following the trend of
 charm decays it dropped to 7.8\% whereas the $\sigma\pi^+$ dominates with 46\%.
 E791 perform a series of alternative fits and tests  to be sure that no 
 other model would as well describe the data.
 
\begin{figure}[t]
 \vspace{7.0cm}
\includegraphics{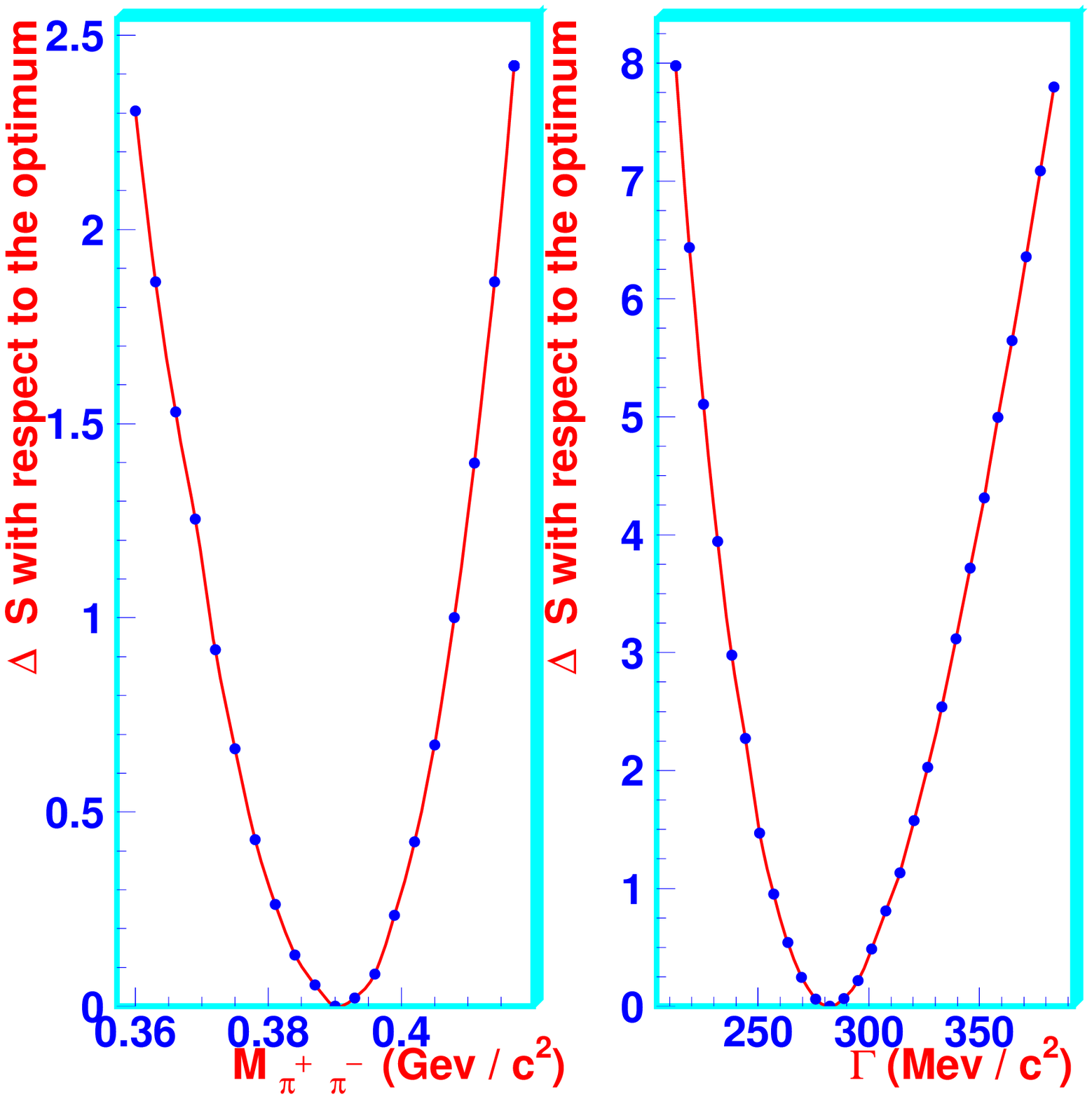}
\includegraphics{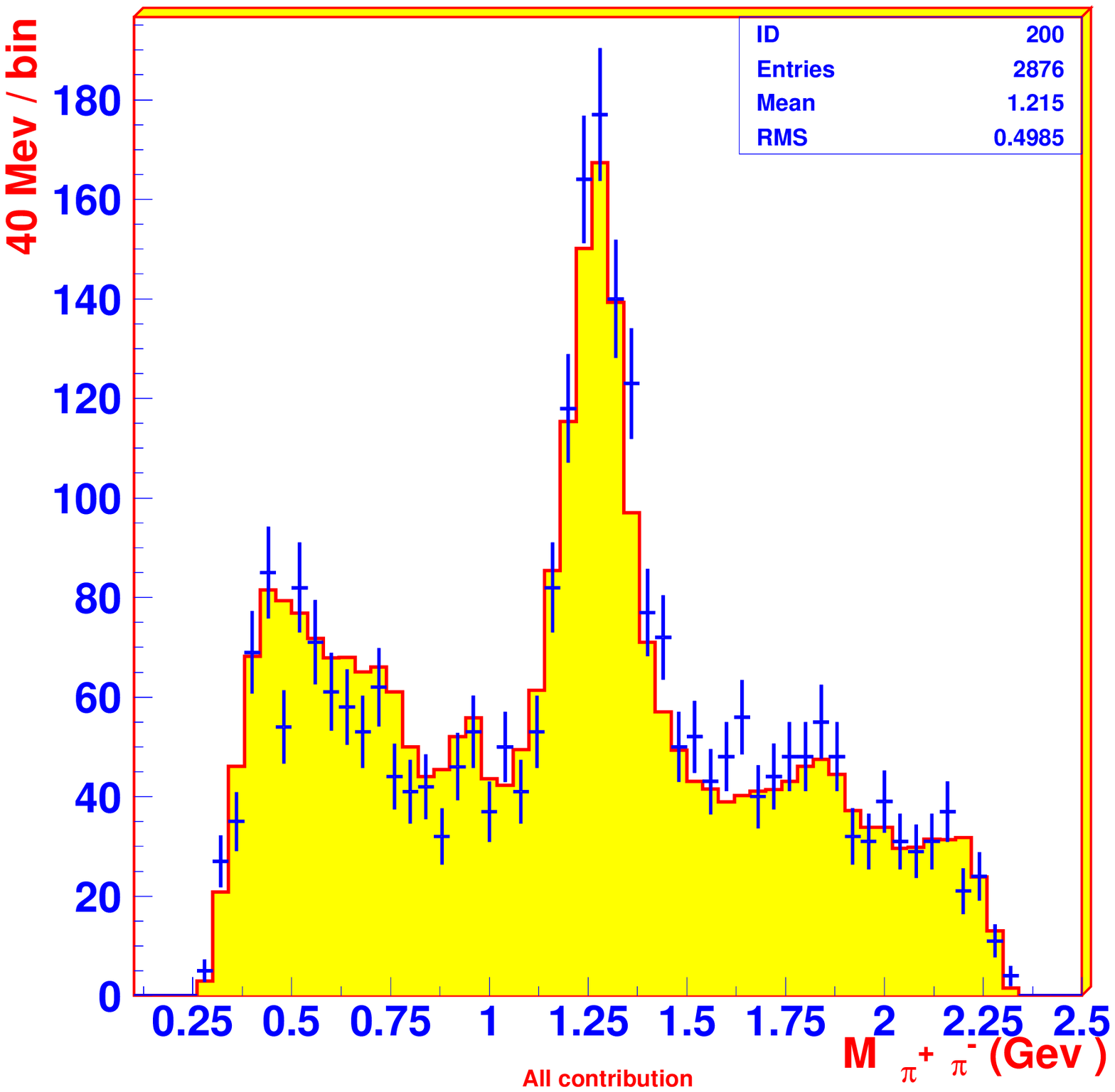}
 \caption[]{\it LEFT: Mass scan on $\sigma$; MIDDLE:width scan 
on $\sigma$; RIGHT: final global fit (error bar is real data 
and histogram is fit )
    \label{fig5} }
\end{figure}

 Studying the channel $J/\Psi \to w \pi^+\pi^-$ BES experiment observe a signal
 of the mesons $\sigma$ and measure their parameters \cite{bes2}; 
 $m_{\sigma} =
490^{+60}_{-36}$ MeV/c$^2$ and $\Gamma_{\sigma} = 282^{+77}_{-50}$ MeV/c$^2$. In figure
 \ref{fig5} we show their final fit and scans for the $\sigma$  mass and  
 width measurements.

\section{The $\kappa$}

\begin{figure}[t]
 \vspace{7.0cm}
\includegraphics{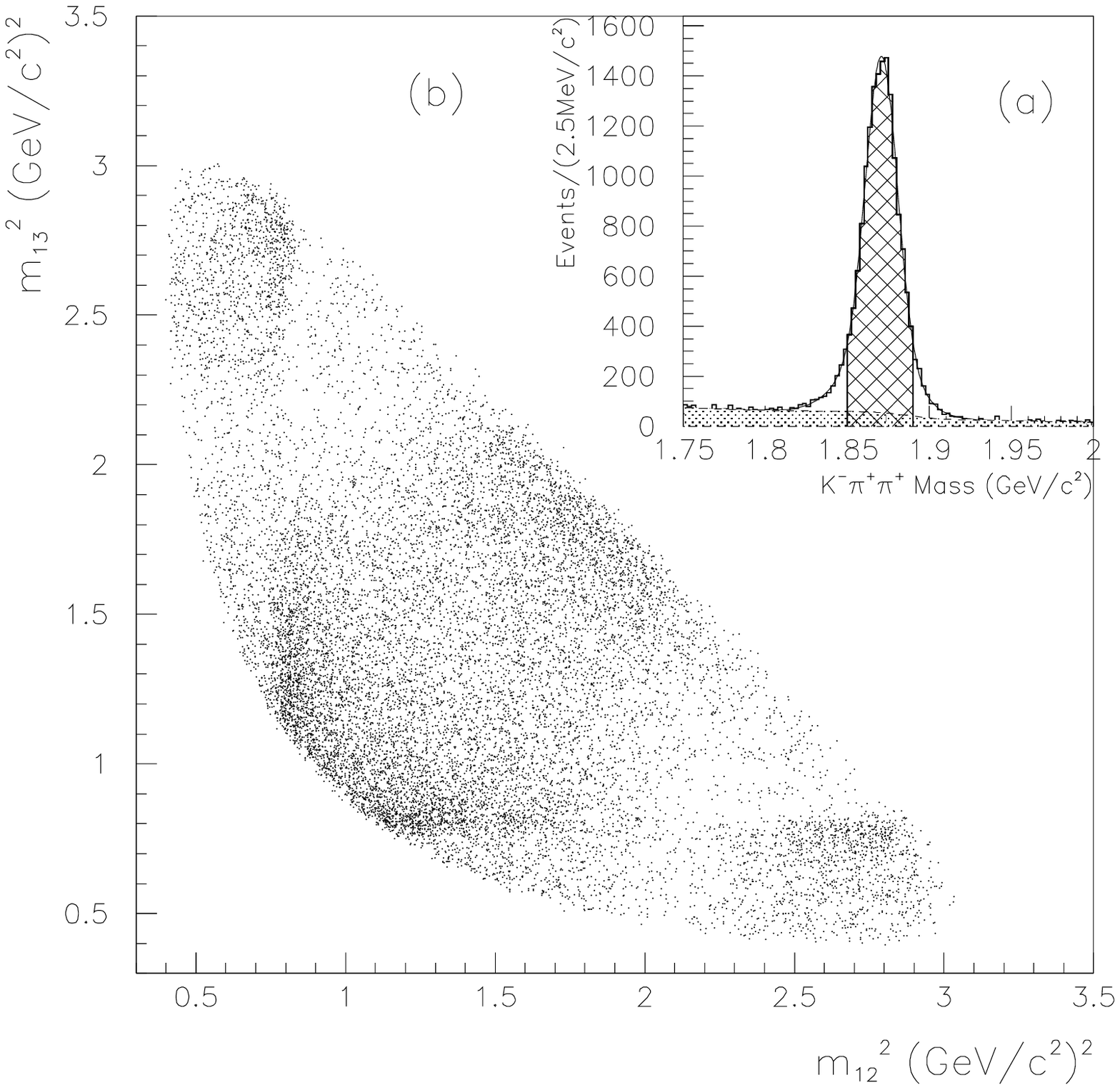}
\includegraphics{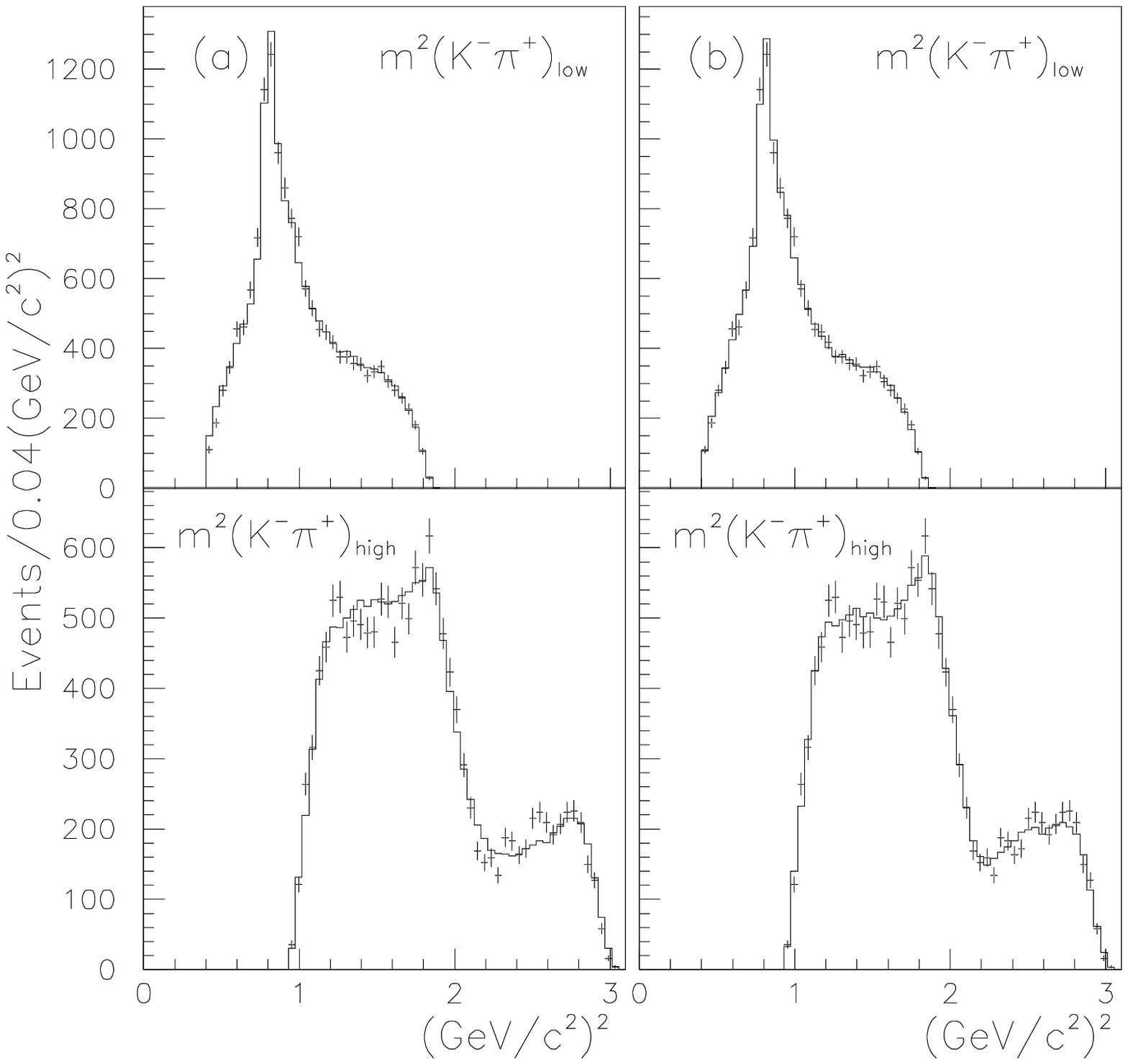}
 \caption[]{\it LEFT: E791 $D^+\to K^-\pi^+\pi^+$ Dalitz plot ; RIGHT:
 projections for  data (error bars)  a)without $\kappa$
   b)with $\kappa$
    \label{fig6} }
\end{figure}

As a last result we consider the $\kappa$ resonance observed in E791 Cabibbo
favored decay $D^+\to K^-\pi^+\pi^+$. Despite the large statistics available in
this channel, no previous experiment \cite{kpipi2}\cite{kpipi3} have been able
to provide a convincing explanation for this decay. E687\cite{kpipi3} best model for their
sample of almost 9000 events have  $\chi^2/dof $ = 87/29. The solution have the
 NR contribution dominating with  a fraction of 99\%.
Large interference  pattern is  produced with all fractions summing 
147\%. In contrast to the $\sigma$  where a clear bump is seen in the
$m^2_{\pi^+\pi^-}$ projection, no evidence for a missing piece in the low
$m^2_{K^-\pi^+}$ region can be easily observed (figure \ref{fig6}b). On the other
hand the very large statistics and small number of possible intermediate 
 states  provide strong evidence of the need for an extra low mass 
 wide resonance contributing for the decay.

The first approach tried by E791 \cite{kpipi} was to include all established
states; ${\bar K}^*(892)\pi^+,  {\bar K}^*_0(1430)\pi^+, 
{\bar K}^*_2(1430)\pi^+,{\bar K}^*(1680)\pi^+ $ plus a NR (they have studied
also the possibility of a non-flat NR contribution without success). In general
the result agree with previous studies including the bad quality of the
fit. At this point they measured mass and width of the scalar $K^*_0(1430)$ to be
1416$\pm$27 and 250 $\pm$ 21 MeV/c$^2$ respectively, which agree with PDG
values. 

Next they include an additional scalar named ``$\kappa$'' 
for which they measure respectively mass and width of 797$\pm$ 47 
and 410$\pm$ 97 MeV/c$^2$. In the same fit the parameters relative to 
$K^*_0(1430)$ were measured to be  1459$\pm$9 and 175$\pm$17 MeV/c$^2$. This new model
describes very well the data with a $\chi^2/dof $ = 46/63, confidence level of
95\%. The contribution of
the new state is dominant with 48$\pm$ 12\% of the total fraction and the NR
contribution is of 13$\pm$7 \%. The sum of all fractions dropped from 134\% in
the fit without $\kappa$ to 88\% indicating a smaller degree of interferences.

\begin{figure}[t]
 \vspace{9.0cm}
\includegraphics{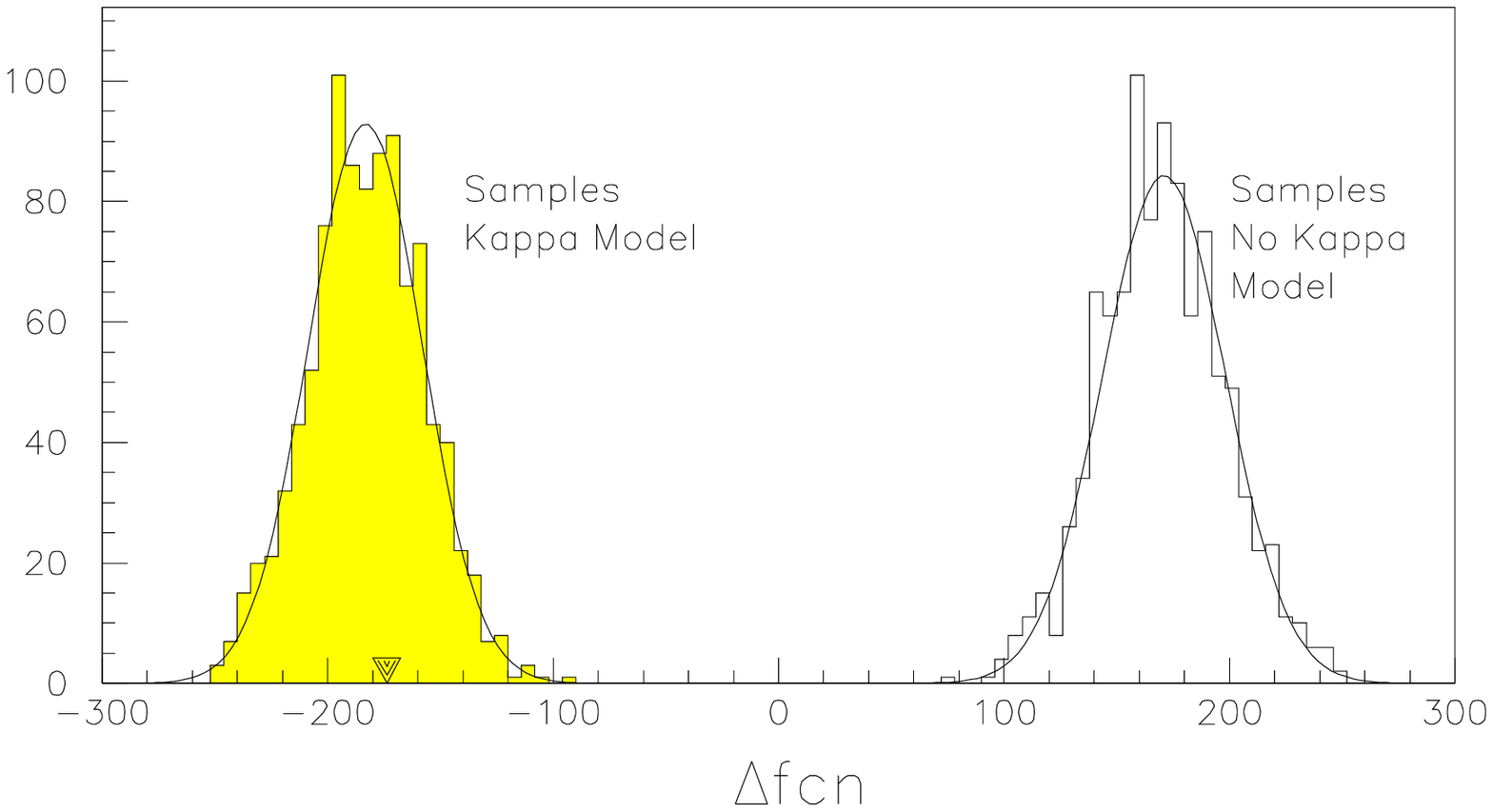}
\includegraphics{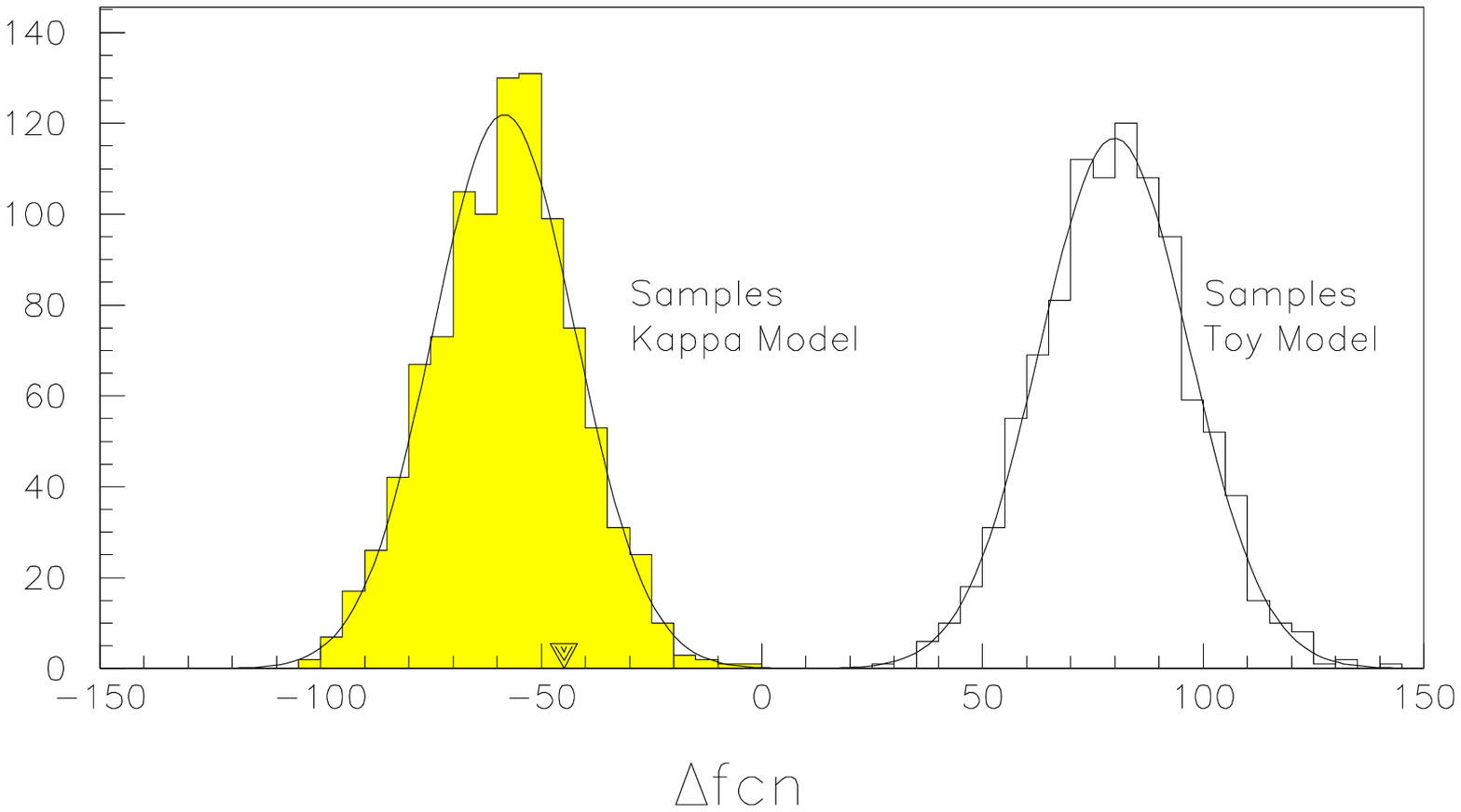}
\includegraphics{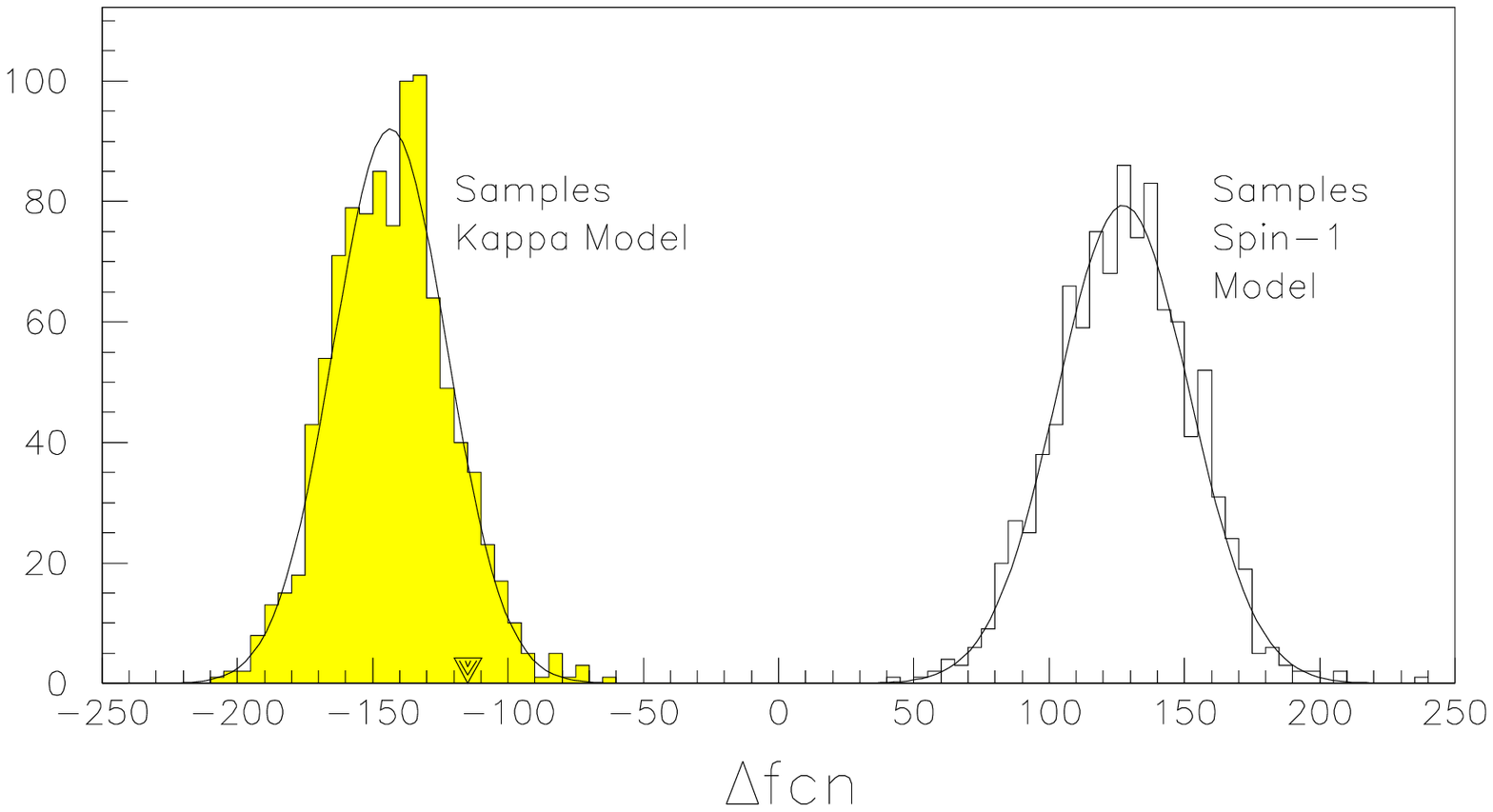}
 \caption[]{\it Comparison of several models to fit E791 $D^+\to K^-\pi^+\pi^+$
 data. Histogram of $\Delta w_{A,B} $, described in the text, for ensembles of
 Monte Carlo ``experiments'' and the data point, solid triangle.
    \label{fig7} }
\end{figure}

In the amplitude analysis described here one do not measure directly a
Breit-Wigner phase, instead it is assumed and consistency tests have to be
performed to
verify if alternative models are able to represent the data. This was 
done for this analysis; a toy-model (consisting of a Breit-Wigner
amplitude without a phase variation), a vector and a tensor alternative models
were tried, none producing a satisfying solutions. When comparing two possible
models $A$ and $B$ the best figure of merit to be used is given by the
 Neyman-Person\cite{np} lema: $\Delta w_{A,B} \equiv -2(ln{\cal L}_A - ln{\cal L}_B)$,
 where  ${\cal L}_{A,B}$ are the likelihood for a given set of events calculated
 with the parameters obtained from fits to the data with models $A$ and $B$. 
 Ensembles of 1000  Monte Carlo ``experiments''
 were generated with the parameters extracted from fits to the data for the two
 models $A$ and $B$. For each of such experiments the quantity $\Delta w_{A,B}$
 was calculated and plotted in figure \ref{fig7}, where the model with the scalar
 $\kappa$ is compared with model without it; with the toy-model $\kappa$ or with
 the vectorial $\kappa$. The discriminating power of the exercise is obvious by
 the separation of the distributions. The value of  $\Delta w_{A,B}$ for the data
 is signed by the triangle showing the clear preference of the data to the model
 of the scalar $\kappa$.

\section{ Conclusion}
 
In conclusion, charm decay is, as said by F.Close and  T$\ddot{\rm o}$rnqvist, a new
window for studying light mesons. We discussed  some  of its remarkable 
contributions. With the already available data sets and the richness of charm
decays, we hope to see in the near
future confirmations and better measurements for some of the resonances, like
the $\kappa$.

\end{document}